\documentclass[%
 reprint,
superscriptaddress,
 amsmath,amssymb,
 aps,
prb,
]{revtex4-1}

\usepackage{amssymb}
\usepackage{fullpage}
\usepackage{graphicx}
\usepackage{bibunits}
\usepackage{wrapfig}
\usepackage{floatflt}
\usepackage{color}
\usepackage{xspace}
\usepackage{dsfont}
\usepackage[official,right]{eurosym}
\usepackage[top=0.8in, bottom=0.75in, left=0.75in, right=0.75in]{geometry}

\usepackage{graphicx}
\usepackage{dcolumn}
\usepackage{bm}
\usepackage{hyperref}

\usepackage{multirow}

\newcommand{\DIMPY} {(C$_7$H$_{10}$N)$_2$CuBr$_4$\xspace}
\newcommand{\BPCB} {(C$_5$H$_{12}$N)$_2$CuBr$_4$\xspace}

\newcommand{\FAPChemical}{(C$_5$H$_6$N$_2$F)$_2$CuCl$_4$\xspace}
\newcommand{\FAP}{(5FAP)$_2$CuCl$_4$\xspace}

\newcommand{\IPA}{(CH$_3$)$_2$CHNH$_3$CuCl$_3$\xspace}
\newcommand{\PHCC}{(C$_4$H$_{12}$N$_2$)Cu$_2$Cl$_6$\xspace}

\newcommand{\be}{\begin{equation} }
\newcommand{\ee}{\end{equation} }
\newcommand{\bea}{\begin{eqnarray} }
\newcommand{\eea}{\end{eqnarray} }

\newcommand{\mb}[1]{\mathbf{#1}}
\newcommand{\mr}[1]{\mathrm{#1}}

\def\XXint#1#2#3{{\setbox0=\hbox{$#1{#2#3}{\int}$}
     \vcenter{\hbox{$#2#3$}}\kern-.5\wd0}}

\begin{document}

\title{Dynamics and field-induced order in the layered spin $S=1/2$ dimer system \FAPChemical}

\author{D. Blosser}
\email{dblosser@phys.ethz.ch}
\affiliation{Laboratory for Solid State Physics, ETH Z\"urich, 8093 Z\"urich, Switzerland}

\author{M. Horvati\'c}
\affiliation{Laboratoire National des Champs Magn\'etiques Intenses, LNCMI-CNRS (UPR3228), EMFL, UGA, UPS, and INSA, Bo\^ite Postale 166, 38042, Grenoble Cedex 9, France}

\author{R. Bewley}
\affiliation{ISIS Facility, Rutherford Appleton Laboratory, Chilton, Didcot, Oxon OX11 0QX, United Kingdom}

\author{S. Gvasaliya}
\affiliation{Laboratory for Solid State Physics, ETH Z\"urich, 8093 Z\"urich, Switzerland}

\author{A. Zheludev}
\email{zhelud@ethz.ch}
\homepage{http://www.neutron.ethz.ch/}
\affiliation{Laboratory for Solid State Physics, ETH Z\"urich, 8093 Z\"urich, Switzerland}

\date{\today}

\begin{abstract}
The quasi-two-dimensional Heisenberg spin $S=1/2$ dimer system bis(2-amino-5-fluoro-pyridinium) tetrachlorocuprate(II) is studied by means of inelastic neutron scattering, calorimetry and nuclear magnetic resonance (NMR) experiments. In the absence of an applied magnetic field we find dispersive triplet excitations with a spin gap of $\Delta=1.112(15)$~meV and a bandwidth of $0.715(15)$~meV within the layers and $0.116(15)$~meV between the layers, respectively. In an applied magnetic field of $\mu_0H_c\approx 8.5$~T the spin gap is closed and we find a field induced antiferromagnetically ordered phase.
\end{abstract}

\pacs{}

\maketitle

\section{Introduction}
Organometallic transition-metal halogen salts have provided a great variety of interesting magnetic model systems. In these compounds magnetic transition-metal ions are linked by halogen mediated super-exchange bridges. Due to the large size and magnetic inertness of the organic cations, exceptionally clean quasi-low-dimensional magnetic systems have been realized in such compounds. Furthermore, the magnetic exchange energy is typically on the order of 1~meV, i.e. $\sim$10~K or $\sim$10~T in units of temperature or magnetic field, respectively, allowing experimental studies of these compounds in their full temperature and magnetic field phase diagram.

Very prominent quasi-one-dimensional organometallic model compounds include the prototypical strong-rung and strong-leg spin ladders \BPCB (BPCB) \cite{Patyal1990, Ruegg2008BPCBThermodynamics, Savici2009, ThielemannRuegg2009, Klanjsec2008NMROrdering, Blosser2017} and \DIMPY (DIMPY) \cite{Shapiro2007, THong2010, Schmidiger2012, Schmidiger2013,Jeong2013, Povarov2015,Jeong2016}, respectively; or the ferromagnetic rung spin ladder \IPA (IPA-CuCl$_3$)\cite{Manaka1997,Masuda2006,Zheludev2007}. 

Finding similar model compounds of quasi-two-dimensional spin systems would be very exciting for the experimental study of such systems and significant effort has been focused on this quest. 
One notable example of a quasi-two-dimensional spin system among the organometallic transition-metal halogen salts is \PHCC (PHCC)\cite{Battglia1988, Stone2001, Perren2015,Bettler2017} which features a rather complicated and partially frustrated quasi-two-dimensional network of spin dimers.
Another whole family of compounds that has attracted attention as candidates for quasi-two-dimensional magnetic model compounds is (py)$_2$CuHa$_4$, where py stands for a pyridine based cation and Ha for a halogen\cite{Place1987, Woodward2002, Hammar2002, Coomer2007, Li2007,Gale2013, Krasinski2017}.

Here, we report on a detailed experimental investigation of the compound bis(2-amino-5-fluoropyridinium) tetrachlorocuprate(II), \FAPChemical or \FAP for short\cite{Li2007}. In this compound, the magnetic Cu$^{2+}$ ions are linked by Cu-Cl$\dotsb$Cl-Cu superexchange bridges into a layered network of dimers\cite{Hong2011}. By means of inelastic neutron scattering experiments we precisely quantify the magnetic exchange interactions and we find non-negligible interactions in all three crystallographic directions. In an applied magnetic field, we find a field induced magnetically ordered phase (BEC of magnons) which we characterize by measurements of specific heat and nuclear magnetic resonance.

\section{Experimental}

\subsection{Crystal structure and magnetic exchange pathways}
\begin{figure}
\includegraphics{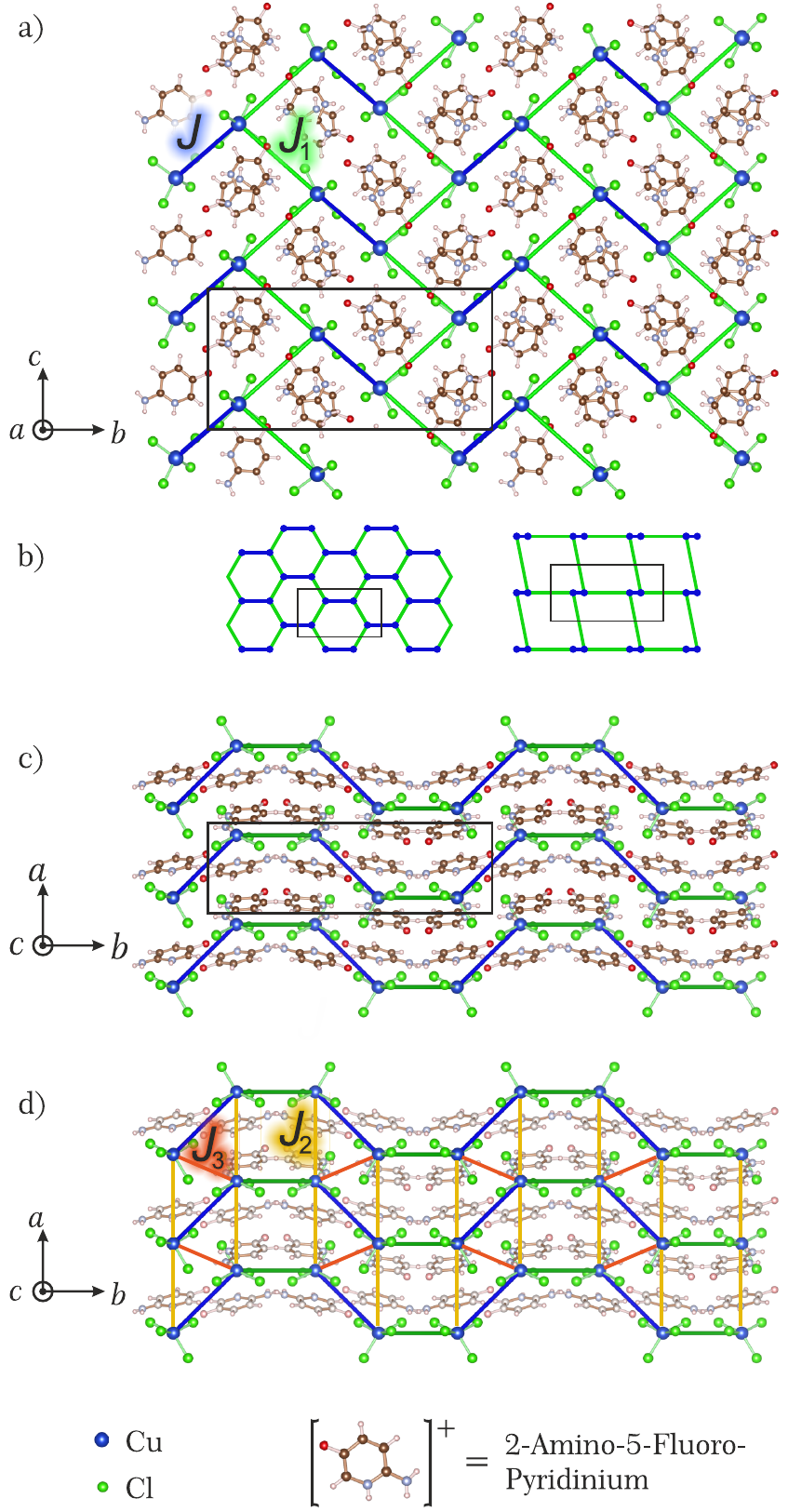}
\caption{\label{fig:structure} 
(a) Crystal structure of \FAP projected along the crystallographic $a$ direction. The blue lines show the dominant magnetic interaction $J$ forming spin dimers. These are coupled in the $b,c$ planes ($J_1$, green lines) creating 2d layers of coupled dimers. These are topologically equivalent to a honeycomb lattice or to a square lattice of dimers as sketched in b).\\
(c,d) Crystal structure as projected along $c$ illustrating the buckled layers. Possible inter-layer interactions are marked by orange and red lines. }
\end{figure}

\FAP crystallizes in a monoclinic $P2_1/c$ structure with lattice constants $a=6.926(7)$~\AA{}, $b=21.73(2)$~\AA{}, $c=10.911(10)$~\AA{} and $\beta=100.19^\circ$ \cite{Li2007}. There are four Cu$^{2+}$ ions per unit cell. These are arranged in buckled [CuCl$_4$]$^{2-}$ anion layers and [C$_5$H$_6$N$_2$F]$^{+}$ cation layers stacked along the crystallographic $a$ direction as shown in Fig. \ref{fig:structure}\footnote{The crystal structure plots shown in Fig.~\ref{fig:structure} are based on renderings generated by the VESTA software package\cite{VESTA2011}.}. 

The dominant magnetic exchange interaction $J$ shown in Fig. \ref{fig:structure} as blue solid lines connects pairs of Cu$^{2+}$ ions to form spin dimers. The weaker interaction $J_1$ links the dimers in the $(b,c)$ planes. These layers are topologically equivalent to a square lattice of weakly coupled dimers, or to a honeycomb lattice as sketched in Fig. \ref{fig:structure}b). 
The stacking of the layers in the crystal structure is shown in Fig.~\ref{fig:structure}c). Possible magnetic interaction pathways between the layers $J_2, J_3$ are indicated in Fig \ref{fig:structure}d)
\footnote{In Ref. \onlinecite{Hong2011}, for the inter-layer coupling, erroneously only $J_2$ had been considered which was argued to be exponentially weaker than the in-plane interactions as it has a much larger Cl-Cl separation (5.07~\AA{}) than the in-plane exchange pathways (3.66~\AA{} and 4.07~\AA{}). However, also considering $J_3$ for which there are two possible exchange pathways with a Cl-Cl separation of 4.24~\AA{} and 4.45~\AA{}, we certainly cannot expect the interlayer interactions to be negligible.}.

\subsection{Synthesis and crystal growth}
For all our measurements we require single crystal samples. Furthermore, in neutron scattering experiments hydrogen leads to very strong incoherent scattering (background in our experiments) necessitating the use of deuterated 2-amino-5-fluoropyridine (d5FAP). This is obtained from protonated 5FAP by catalytic H/D substitution\cite{Esaki2006,Atzrodt2007} as follows: 4~g of 5FAP is slurried in 62.3~ml of D$_2$O in a PTFE lined 125~ml pressure vessel with 200~mg of Pd/C (10 wt.\%) catalyst. The pressure vessel is closed under Ar atmosphere with 10~ml of H$_2$ gas. For the substitution reaction to occur, the vessel is kept at 180$^\circ$C for 72~h whilst continuously stirring. Finally, after filtering off the catalyst and drying the product, $\sim$70\% deuterated 2-amino-5-fluoropyridine is obtained as confirmed by mass spectroscopy.

Single crystals are grown by slowly cooling a saturated solution: d5FAP (21~mmol) and CuCl$_2$ (10.5~mmol) are dissolved in 15~ml D$_2$O and 5.4~ml of DCl in D$_2$O (35\% wt.)\cite{Li2007}. The obtained solution is evaporated to obtain approx. 10~ml of saturated solution at 15$^\circ$C. Small seed crystals are hanged into the solution on a thin PTFE thread. Slowly cooling the saturated solution from 15$^\circ$C to 2$^\circ$C at a rate of 0.5~K/day large green high quality single crystals of $m\approx 0.4$~g are obtained.

\begin{figure*}[t]
\includegraphics{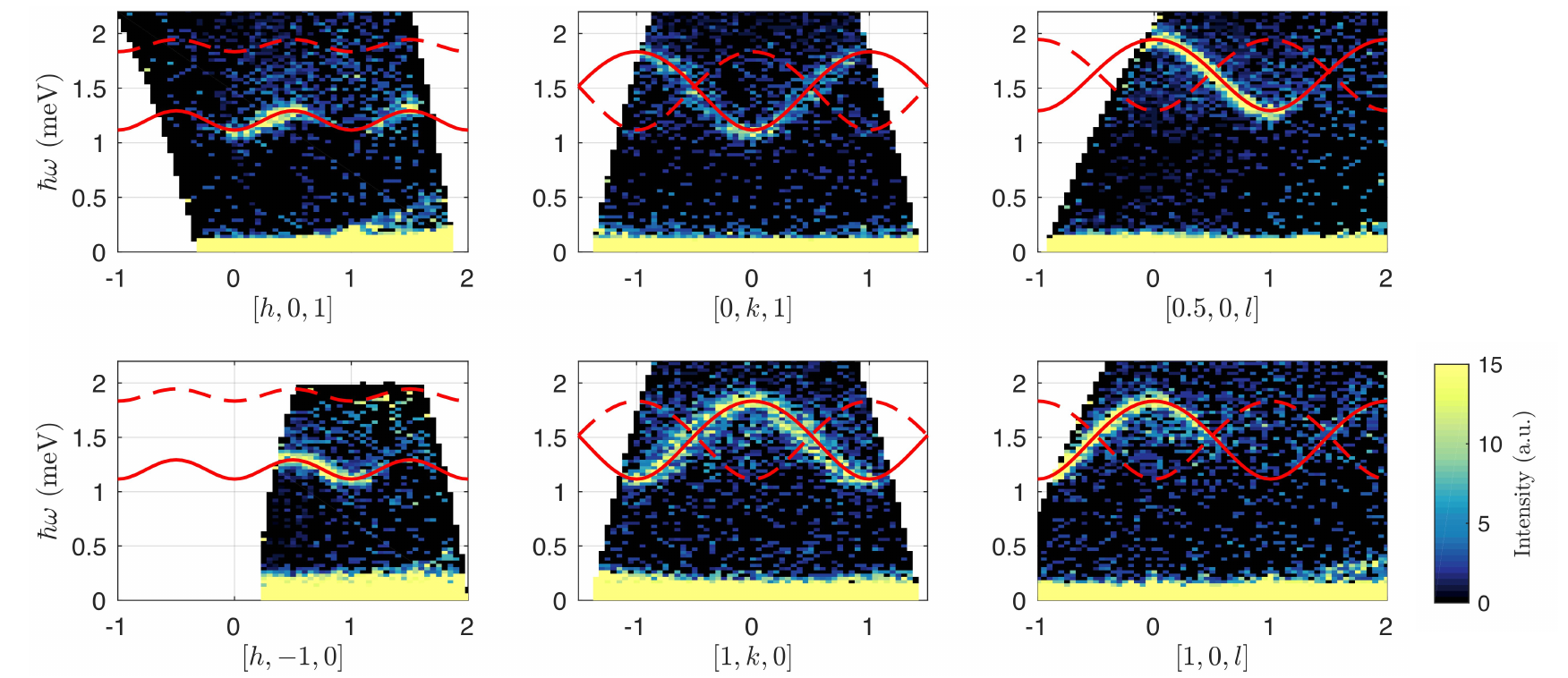}
\caption{\label{fig:neutrons}Representative false-color slices through the 4-dimensional inelastic neutron scattering data set measured in \FAP at $T=100$~mK in zero magnetic field showing the dispersion of magnetic triplet excitations. The solid and dashed lines correspond to the calculated triplet dispersion branches as described in the text. }
\end{figure*}

\subsection{Experimental Methods}
All our experiments are performed {on $\sim 70$~\% deuterated} single crystal samples with the applied magnetic field aligned with the crystallographic $b$ direction 
{\footnote{{Additional heat capacity data was measured on a fully protonated crystal and the results were virtually identical with the data obtained on $\sim 70$~\% deuterated samples. Also, the lattice constants of the crystal do not change measurably upon deuteration. For this reason we do not expect any significant difference between protonated and deuterated samples. Similarly, we do not expect the partial deuteration to cause any `disorder' in regard to the sample's magnetic properties.}}}. 

\paragraph{Inelastic Neutron Scattering.}
For the neutron scattering experiments 6 crystals (approx. 70\% deuterated) were wrapped in a thin PTFE film\footnote{It was found that \FAP chemically reacts with aluminium on a time scale of a few weeks.}, co-aligned to within 1.5$^\circ$ and fixed on an aluminium sample holder. The neutron scattering experiments where performed at the LET time-of-flight spectrometer\cite{Bewley2011} at the ISIS facility, UK. The sample was mounted on a dilution refrigerator in a 9~Tesla cryomagnet with the crystallographic $b$ direction parallel to the magnetic field, i.e. perpendicular to the scattering plane. Making use of repetition rate multiplication, data were simultaneously collected using $E_i=2.1$, $3.5$ and $6.9$~meV incident energy neutrons. The data\cite{DataISIS} has been analyzed using the HORACE software package\cite{Horace}.

\paragraph{Specific Heat.}
Specific heat was measured on a $m=1.15(8)$~mg single crystal using a Quantum Design PPMS equipped with a $^3$He-$^4$He dilution refrigerator insert.

\paragraph{Nuclear Magnetic Resonance.}
$^{19}$F (nuclear spin $I=1/2$) NMR experiments were performed on a single crystal of dimension $\sim3.6\times1.2\times0.9$~mm with an rf coil precisely fitting the sample dimension. This assembly is mounted inside the mixing chamber of a $^3$He-$^4$He dilution refrigerator ensuring good thermalization. The external magnetic field is applied parallel to the crystallographic $b$ direction. 
The nuclear relaxation rate $1/T_1$ is obtained from the spin-echo intensity $M(t)$ measured as a function of time $t$ after a saturating pulse by fitting the exponential function $M(t)=M_\text{eq}-M_0\exp \{-(t/T_1)\}$ to the data where $M_\text{eq}$ is the intensity in thermal equilibrium. These data were measured at the spectral center or at the upper peak when the spectrum is split within the ordered phase. At several points inside the ordered phase, we verified that the thus obtained $T_1$ is identical for both of the split peaks.

\section{Results and Discussion}

\subsection{Inelastic Neutron Scattering}
Neutron scattering intensities were measured at $T=100$~mK in zero field as a function of energy and wave-vector transfer to obtain a large 4-dimensional data set $\mathcal{I}(\mb{Q},\omega)$. Representative cuts through this data are shown in Fig. \ref{fig:neutrons}. The false-color plots show scattering intensity vs. energy and momentum transfer, where the data has been integrated in slices of $\pm0.1$ reciprocal lattice units (r.l.u.) along the reciprocal space directions not shown in the plots.

We find resolution-limited dispersive triplet excitations with a spin gap of $\Delta=1.112(15)$~meV. They show a sizable dispersion in the $k,l$ directions (i.e. within the dimer planes) with a bandwidth of $\Gamma_{k,l}=0.715(15)$~meV. 
The inter-layer dispersion along $h$ is significantly smaller with a bandwidth $\Gamma_h=0.116(15)$~meV.

\begin{figure}[h]
\includegraphics{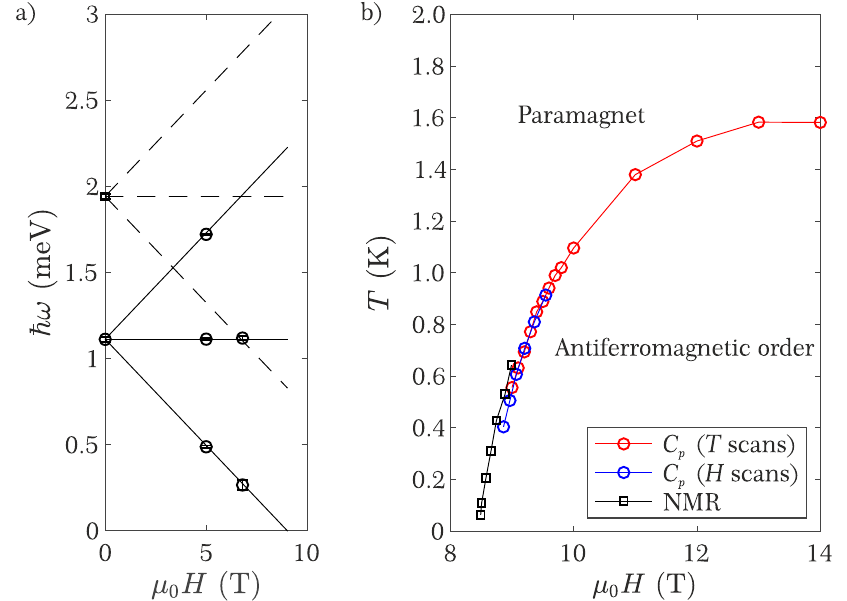}
\caption{\label{fig:NeutronsField}a) Band extrema of the Zeeman-split triplet modes as a function of magnetic field. b) Phase boundary of the field induced long-range ordered antiferromagnetic phase as measured by specific heat and nuclear magnetic resonance.}
\end{figure}

To determine the relevant exchange parameters we extend the procedure described in Ref. \onlinecite{Hong2011} to also include the possible inter-layer interactions. We start by considering a lattice of two isolated dimers per unit cell with an antiferromagnetic exchange interaction $J$. The weaker exchange interactions $J_1,J_2,J_3$ between the dimers we now treat within the random phase approximation (RPA)\cite{Hong2011,JensenMackintosh1991}. For the triplet dispersion we thus find
\begin{align}
(\hbar\omega^{\pm}_\mb{q})^2    =  J^2 \pm J R(T)  & \left[    J_1 \left( \cos \pi (k+l)     +\cos \pi (k-l)    \right)   \right.  \notag  \\
&\,\,\, \left. \pm (J_2-2J_3)\cos(2 \pi h) \right] 
\end{align}
with $R(T)=\frac{1-e^{-J/(k_B T)}}{1+3 e^{-J/(k_B T)}}$ denoting the difference in population of the singlet and triplet states of an isolated dimer. As there are two dimers per unit cell we naturally find two excitation branches. However, in almost all of reciprocal space accessed in our experiments only one excitation branch carries all the intensity.
Fits of this dispersion to the neutron scattering data are shown in Fig.~\ref{fig:neutrons} as solid and dashed lines. The fitted estimates for the exchange parameters are
\begin{align}
J &= 1.586(8)\,\,\mathrm{meV}\notag\\   
J_1 &= 0.333(7)\,\,\mathrm{meV}\\   \notag
(J_2-2J_3) &= -0.139(8)\,\,\mathrm{meV}.   \notag
\end{align}
At the RPA level, the dispersion only contains the sum $(J_2-2J_3)$ of the possible exchange interactions between the planes and we cannot determine $J_2$ and $J_3$ independently. However, the relevant magnetic exchange interactions between the Cu$^{2+}$ ions are mediated by Cu-Cl$\dotsb$Cl-Cu super-exchange pathways leading to antiferromagnetic exchange $(J>0)$. Ferromagnetic interactions $(J<0)$ seem unlikely in this setting\cite{Straatman1984, Landee2013}. Since for $(J_2-2J_3)$ a negative value is found, we expect that the inter-plane interactions are predominantly due to $J_3$. Moreover, the relevant Cl$\dotsb$Cl separation for $J_2$ is larger than for $J_3$, again pointing to $J_3$ as the dominant inter-layer coupling. Assuming $J_2=0$, we find a ratio $J_3/J_1=0.21$ of the inter-layer vs. in-plane coupling strength, quantifying the quasi-two-dimensionality of this compound.

{ In an applied magnetic field, the three-fold degeneracy of the triplet modes is lifted and the two dispersion branches are Zeeman-split into six bands. This we clearly see in neutron scattering data obtained at $\mu_0H=5$ and 6.8~T. From these data sets we extract the band extrema as shown in Fig.~\ref{fig:NeutronsField}a).} 
From the field dependence of the Zeeman shift we can directly determine the $g$-factor in this geometry as $g_b=2.13(2)$. Extrapolating the low energy triplet band minimum we expect the spin gap to close at $\mu_0H\approx 9$~T. Indeed, in an applied magnetic field slightly below this value, at low temperatures we find a phase transition to a magnetically ordered state as revealed by measurements of specific heat and nuclear magnetic resonance.

\begin{figure}[t]
\includegraphics{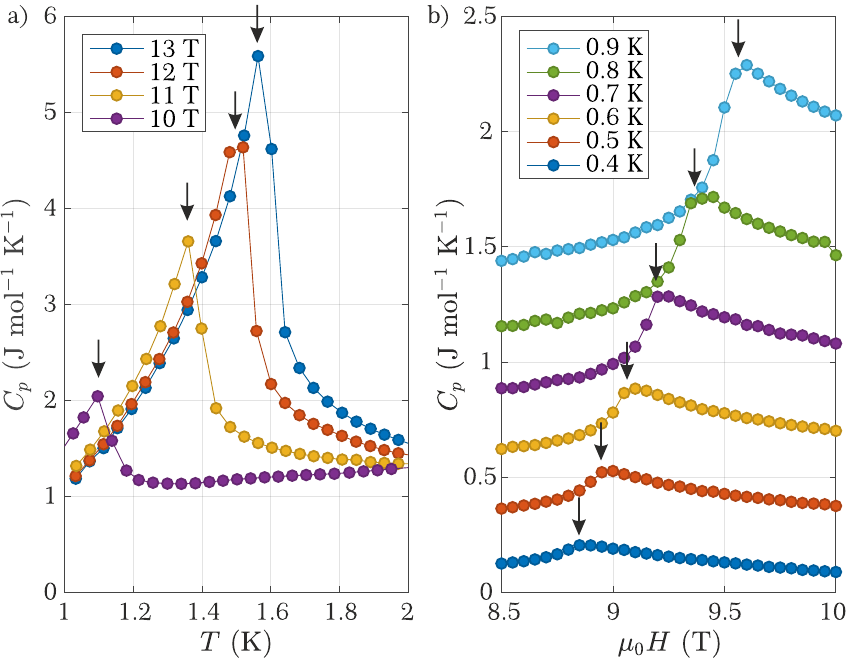}
\caption{\label{fig:SpecificHeat}Representative measurements of specific heat versus temperature (a) and magnetic field (b). Peaks mark the onset of long-range antiferromagnetic order.}
\end{figure}

\subsection{Specific heat}
Specific heat was measured in a wide range of temperatures and magnetic fields up to 14 Tesla. Representative data vs. temperature and magnetic field, respectively, are shown in Fig.~\ref{fig:SpecificHeat}. At high magnetic fields, we find a very strong and very sharp peak in specific heat vs. temperatures (Fig.~\ref{fig:SpecificHeat}a). Towards lower temperatures and closer to the critical field $H_c$, the magnitude of the peak decreases and it broadens. Data traces of specific heat vs. magnetic field measured at lower temperatures (Fig.~\ref{fig:SpecificHeat}b) also reveal a maximum that decreases and broadens towards lower temperatures. Below 0.4~K pinpointing the phase boundary becomes impossible. {The reason for this decrease and broadening of the peaks remains unclear.}

The phase boundary extracted from this data is plotted in Fig.~\ref{fig:NeutronsField}b). It has a maximum near $13.5$~T where it occurs at a temperature of $T_N=1.58$~K. Towards zero temperature the phase boundary extrapolates to a value slightly below 9~T, consistent with the field where the spin gap closes as estimated from the inelastic neutron scattering data. 
{Assuming the phase boundary to show a roughly symmetric `dome'-shape between the lower critical field $H_c$ and the field of saturation $H_\mr{sat}$, we expect an upper critical field of $\mu_0H_\mathrm{sat}\approx$~18--19~T, in line with the saturation field estimated from high field magnetization measurements\cite{Li2007}.}

\begin{figure*}[t]
\includegraphics{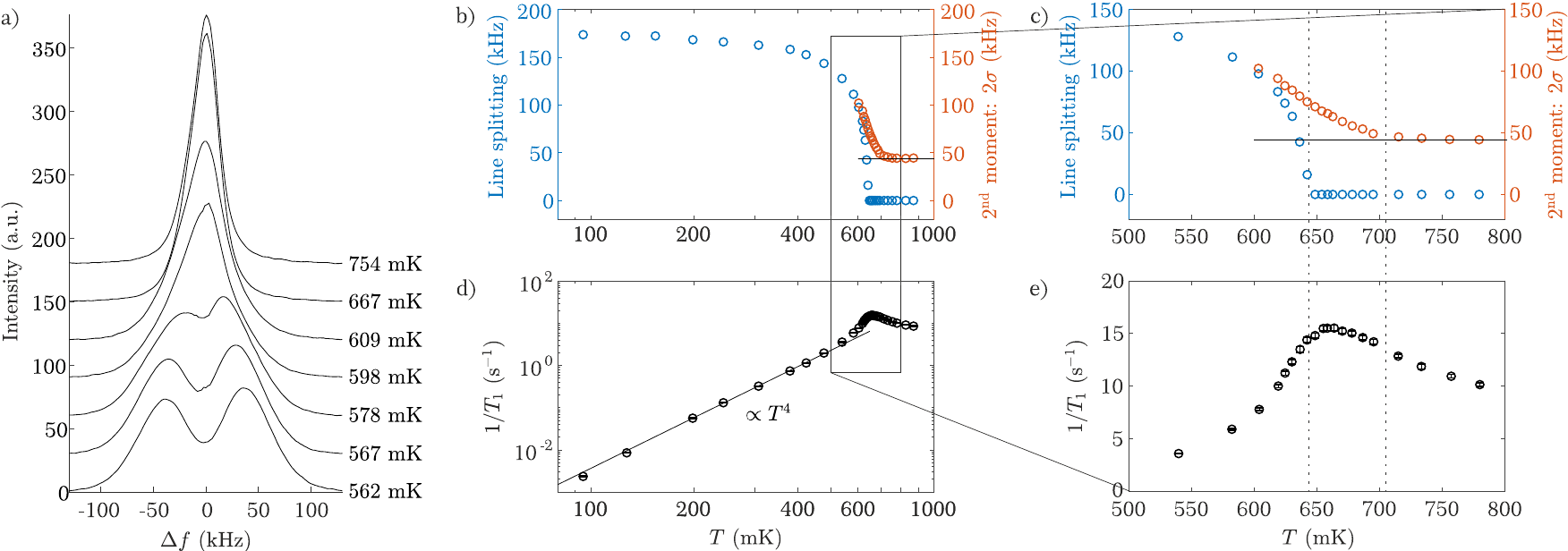}
\caption{\label{fig:NMR1}NMR data measured at $\mu_0H=9$~T. a) Evolution of the $^{19}$F NMR spectrum upon entering the ordered phase. The splitting of the resonance indicates antiferromagnetic long range order. b,c) Temperature evolution of the line splitting and (d,e) the nuclear spin relaxation rate $1/T_1$.}
\vspace{-1mm}
\end{figure*}

\subsection{Nuclear magnetic resonance}
To investigate the nature of the phase transition and of the field induced long-range ordered phase we turn to NMR measurements of the $^{19}$F nuclear spin ($I=1/2$). Measurements of both the NMR spectrum as well as the nuclear spin relaxation rate have been performed at various magnetic fields and temperatures. In the \FAP crystal structure there are 2 inequivalent F sites at low symmetry positions. Nonetheless, the NMR lines from the two sites apparently overlap and therefore could not be distinguished.

For $\mu_0 H=9$~T, representative NMR spectra are depicted in Fig.~\ref{fig:NMR1}a). Whilst we observe a single sharp resonance line at elevated temperatures, upon cooling this line slightly broadens and splits, confirming a transition to an antiferromagnetic state. 

Analyzing the evolution of the spectrum in more detail, in Fig~\ref{fig:NMR1}b,c) we show the apparent line splitting taken as the distance between the peaks, whenever two distinct maxima are visible. This simple measure will slightly underestimate the antiferromagnetic order parameter in the regime where the splitting is of similar size as the width of the individual lines. However, we note that the lines do not only split, but also the lineshape changes across the transition. For this reason, fitting a double peak function to the measured spectra to more accurately extract the line splitting has not proven useful.

A different and well-defined measure of the observed spectra is the second moment shown as $2\sigma$ in Fig~\ref{fig:NMR1}b,c) 
{\footnote{{The second moment $\sigma$ of a peak-shaped function $f(x)$ about its mean $\mu$ is defined as 
$\sigma^2 \!=\! \frac{1}{A} \int_{-\infty}^\infty  \left(x-\mu\right)^2 f(x) \mr{d}x $, 
where
$\mu \!=\! \frac{1}{A} \int_{-\infty}^\infty  x f(x) \mr{d}x $ and  $A \!=\!\int_{-\infty}^\infty f(x) \mr{d}x $.
The second moment provides a measure for the variance of a function. 
If there is a single peak, $\sigma$ measures its width. 
For example, for a single Gaussian peak $g(x)=\frac{1}{c\sqrt{2\pi}}e^{-\frac{1}{2}(x/c)^2}$ we find $\sigma=c$. On the other hand, for two strongly separated peaks, $\sigma$ measures the distance between the peaks. 
For the case of two Gaussians $g(x+L)+g(x-L)$ the second moment is $\sigma=\sqrt{L^2+c^2}$,  which reduces to $\sigma\approx L$ in the limit $L\gg c$. For the data analysis at hand, the second moment is thus a convenient and model independent measure with a transparent interpretation both in the limit where we observe a single peak as well as in the limit where two strongly separated peaks are observed.}}} 
Regardless of the lineshape, for strongly split peaks, $2\sigma$ exactly corresponds to the line splitting. We observe that already at temperatures roughly 50~mK above the appearance of two distinct peaks, the line starts to broaden and its second moment increases. The onset of this line broadening directly reflects the onset of slow fluctuations with a frequency that is lower than the $^{19}$F nuclear spin Larmor frequency of $f_0\sim 360$~MHz at the field of 9~Tesla.

Measurements of the nuclear spin relaxation rate $1/T_1$ at 9~Tesla are plotted in Fig.~\ref{fig:NMR1}d,e). At low temperatures, well inside the ordered phase, we observe a $1/T_1 \propto T^4$ power-law dependence over almost one decade in temperature. The high value of the power-law exponent is only slightly smaller than what is usually observed in BEC phases of quantum antiferromagnets (5.0--5.5)\cite{Mayaffre2000,Jeong2017}, reflecting a high-order relaxation process\cite{Beeman1968}. In the vicinity of the transition, we find a rounded peak with a maximum at slightly higher temperatures than the appearance of the apparent line splitting.

\begin{figure}[]
	\includegraphics[width=0.5\textwidth]{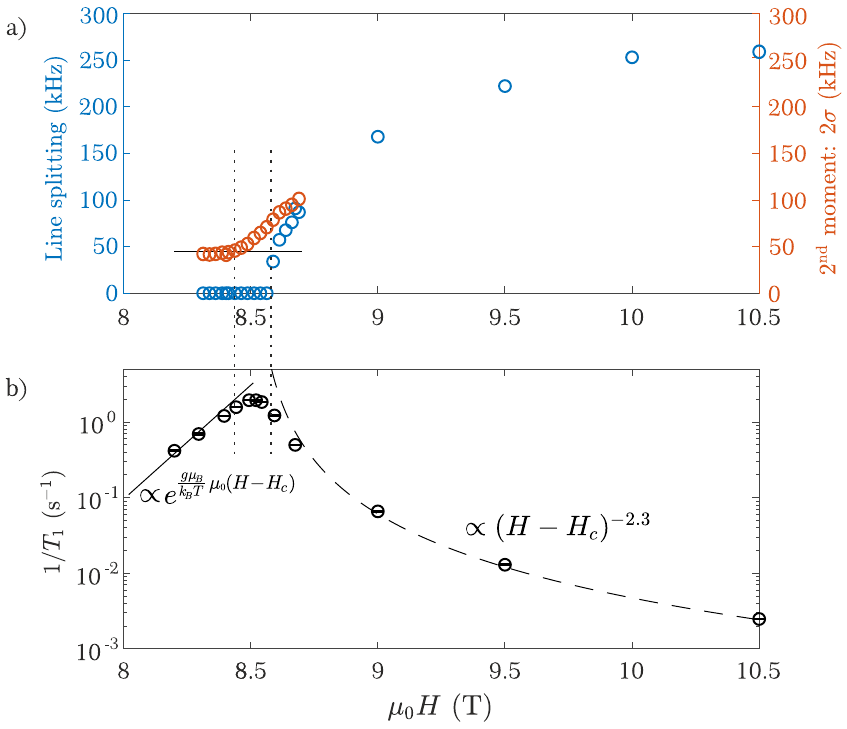}
	\caption{\label{fig:NMR2}a) NMR line broadening and splitting and (b) $1/T_1$ relaxation rate vs. magnetic field at $T=206$~mK.}
	\vspace{-1mm}
\end{figure}

Similar data measured at a constant temperature of $T=206$~mK versus magnetic field is shown in Fig.~\ref{fig:NMR2}. We again observe a broadening of the NMR line (characterized by the second moment) followed by a splitting into two peaks. Again, the phase transition is accompanied by a maximum in $1/T_1$ at $\mu_0 H=8.51$~T, slightly before the line splitting becomes apparent. Taking this value as the critical field $H_c$, at lower fields we expect an exponential reduction of the relaxation rate $1/T_1\propto e^{\frac{g\mu_B}{k_B T}\mu_0(H-H_c)}$, indicative of the opening of a spin gap. (Here, using the previously determined value of $g=2.13$, the only unknown parameter is an overall prefactor.) We only have very few data points below the critical field but they roughly seem to follow the expected trend. Inside the ordered phase, phenomenologically, the data follows a power-law behavior $1/T_1\propto (H-H_c)^{-\kappa}$ with an exponent $\kappa\approx 2.3$. 


We performed spectral NMR measurements at various temperatures and magnetic fields. From these data we have extracted the phase boundary as the point where two distinct maxima become visible in the spectrum. These points are shown in the phase diagram of Fig.~\ref{fig:NeutronsField}b). 
{At low temperatures the phase boundary becomes very steep and seems consistent with a $T_N\propto (H-H_c)^{2/3}$ behavior as has been observed in fully three-dimensional systems \cite{Zapf2014}. However, from the present data it is impossible to reliably extract a critical exponent.}
We observe a slight mismatch between the points obtained from NMR measurements and the position of the phase boundary extracted from specific heat data. Since for these measurements two different samples were used, we suspect this small difference to be due to slightly different transition fields in the two samples, or a slightly different alignment of the crystals with respect to the applied magnetic field in the two experiments.

\section{Conclusion}
The layered spin dimer compound \FAP has been studied in detail. We find a singlet ground state and dispersive triplet excitations separated by a spin gap of $\Delta=1.112(15)$~meV. The triplon dispersion shows a bandwidth of $\Lambda_{k,l}=0.715(15)$~meV within the dimer planes and $\Lambda_{h}=0.116(15)$~meV between the layers.

Measurements of specific heat and nuclear magnetic resonance reveal a field induced antiferromagnetically ordered phase beyond $\mu_0H_c\approx 8.5$~Tesla extending up to 1.58~K at 13.5~T. 
\\

{\it Acknowledgement.}
We would like to thank David Schmidiger for his involvement in the early stages of this project. We thank M. Turnbull (Clark University) for his advice regarding the catalytic deuteration of 2-amino-5-fluoropyridine. This work is partially supported by the Swiss National Science Foundation under Division II.

\bibliography{references5FAP}

\end{document}